\documentclass[pra,twocolumn,floats,floatfix]{revtex4}
\usepackage{graphicx,epsfig}
\usepackage{times}
\usepackage{bm}
\usepackage{amsmath,amssymb}

\begin{document}

\title{Viscous Fluids and Quantum Mechanics}
  \author{Jo{\~a}o Belther Junior}
  \affiliation{Instituto de F{\'i}sica, Universidade de S{\~a}o Paulo,
  Brasil}
\email{belther@fig.if.usp.br}
  \date{\today}

\begin{abstract}
From a simple analysis of particle orbits and fluid flows in presence
or not of dissipation, some connections between apparently
uncorrelated research areas are made. The main results point out for
a deep relation between quantization of classical conservative systems
and the dissipative version of these systems.
\end{abstract}

\maketitle

The aim of this work is to point out that different areas of research
can mutually benefit from each other if an appropriate translation is
made. To make an intelligible translation the fundamental deductions
will be introduced by the using of widely known concepts and the
consequences will be more specific. The so called fundamental concepts
are input and output information, and some evolution (dynamical)
law. For example, with the input information of the postion and
velocity at some instant of a particle obeyng Newton's second law, it is possible to
determine the future positions and velocities for some quantity of
time, in general not infinite, if the force (potential) field
satisfies some (smoothness) conditions. If the input information is,
for example, the positions in two different instants, it is not
possible in general to assign a definite orbit as output
information. For periodic systems, even with a huge ammount of
positions in different instants as input, it is not possible to decide
the output orbit, if the positions and instants been measured are rational
numbers, as is ever the case for (finite precision=) physical
measurements. The very dynamical law can possess some undecidibility,
as is the case for singularities in the force field or if one tries to
describe physical motion of a particle by first-order autonomous ordinary
differential equations in position only: the simple one-dimensional harmonic
oscillator would present some problems, for would exist two velocities
associated to the the same point. This latter description is very
important for conservative forces, for in this case the velocity in
cartesian coordinates is equivalent to the gradient of the action
function, solution of the Hamilton-Jacobi equation. This point will be
made clear in the sequence, but before some important equations will
be deduced.

If the orbit is sufficient smooth in some region, it is possible to
consider it as the integral curve for some vector field
\begin{equation}
\dot{\mathbf{x}(t)}=\mathbf{v}(\mathbf{x}(t),t).
\end{equation}
For a conservative force field, where the force can be written as
(minus) the gradient of some potential $V(\mathbf{x}(t),t)$, a time
differentiation of the last equation leads to
\begin{equation}
\frac{\partial\mathbf{v}}{\partial
  t}+\mathbf{v}\cdot\nabla\mathbf{v}=-\nabla{V},
\end{equation}
wich is the Euler equation of fluid dynamics, if a identification
between the potential and the pression $p(\mathbf{x},t)$ is made.
The Euler equation has the additional requirement of
incompressibility, $\nabla\cdot\mathbf{v}=0$, and appropriate boundary
conditions. It is simple to show that for a gradient velocity field,
$\mathbf{v}=\nabla S$, the Euler equation reduces to the
Hamilton-Jacobi (HJ) one
\begin{equation}
\frac{\partial S}{\partial t} +\frac{1}{2}\nabla S\cdot\nabla S +V=0.
\end{equation} 
Altough the equations (dynamical law) are formally the same, there is a fundamental
difference in their commom use: for the HJ equation, the potential
(pression) is given, it is part of the input information, and the
incompressibility condition is not imposed. In fact, apparently there
is no study even if the HJ equation admit harmonic solutions for
general boundary conditions. As is widely known, the name action
function is ascribed for the relation between the solution of HJ
equation and the action, generally given by the functional of the
Lagrangian. Here is the moment for another observation. It is
generally said that the minimum action principle is equivalent to the
Newton's second law. This is true if the velocity and position are
given at the same instant in the two descriptions. Strictily, the
minimum action is stated for two different points in space-time, the
velocity should be considered as additional input information. For
very near instants and positions it can be said the two descriptions
are equivalent, but it is important to emphasize this local character
for the equivalence. The problem
is analogous to that of geodesics for a given metric: for very near
points, the geodesics are in fact the minimum distance curves
connecting this points, but this cannot be extended to arbitrary
points. For fixed energy $E$, in a gradient force field, the orbits are the geodesics of the
Maupertois metric: $g_{ij}=(E-V)\delta_{ij}$. It is a simple task, not
generally made, to calculate some geometrical quantities from this
metric, like an invariant  density, a connection
compatible with the metric and the curvature. This geometrical
analysis will be treated after the introduction of dissipation.

The Euler equation is a good tool for describing long wavelength (low
frequency) phenomena, where the effects of turbulence, a short-range 
viscosity dependent phenomenon, can be neglected. It is the case for
the waves in the ocean or in a pond, where the wavelenghts varies from
fractions of a meter to several hundred meters. If one considers the
photon fluid (gas) of blackbody radiation, for low frequency the
classical prediction for the energy density is in good agreement with
experimental data, but not for very higher ones, when quantum
mechanics turns out to be a necessity. In the sequence will be
explored the possibility to explain these ``quantum effects'' in the
photon fluid, or possibly any other system, by a similar approach as
for the ocean waves: considering dissipation in the system. For the
photon fluid is something like to suppose a dissipation coefficient so
small that for low frequencies it can be, for almost every purpose, be
ignored, but not for very high frequencies. As is generally known, the
HJ equation is in some sense the ``classical'' version of the
Schr{\"o}dinger equation, it was from the HJ equation that
Schr{\"o}dinger obtained his wave equation, in a heuristic
derivation. What is not apparently explored is the fact that
Schr{\"o}dinger equation is the Navier-Stokes (NS) equations, if a
irrotational (gradient) velocity field is considered and a appropriate
simple transformation is made. One can readly 
see this by considering the wave equation for unity mass and imaginary
time, $t\rightarrow -it$
\begin{equation}
 \hbar\frac{\partial\Psi}{\partial
   t}+V\Psi-\frac{\hbar^2}{2}\nabla^2\Psi=0,
\end{equation}
and making the transformation $\Psi=e^{\frac{S}{\hbar}}$ to obtain
\begin{equation}
\frac{\partial S}{\partial t}+\frac{1}{2}\nabla S\cdot\nabla S
-\frac{\hbar}{2}\nabla^2 S=0,
\end{equation}
wich is the NS equation 
\begin{equation}
\frac{\partial\mathbf{v}}{\partial
  t}+\mathbf{v}\cdot\nabla\mathbf{v}+\nabla p -\nabla^2\mathbf{v}=0,
\end{equation}
for an irrotational velocity field. It is important to notice that if
the condition of incompressibility is imposed on $\mathbf{v}$ for NS
equation with a gradient field, the NS equation reduces to the Euler
equation, except for the boundary conditions. In a analogous situation
encountered before, it is important to remark that apart the formal
equality between the equations, there is the fundamental difference
that in Schr{\"o}dinger equation the potential is generally given. By
this way, one can consider the (irrotational) NS equation as a
Schr{\"o}dinger equation where the potential is not given from the
start, it is not part of the input information. Apart from the connection with the NS equation, this kind of
analysis for the Schr{\"odinger} equation without a definite potential
is encountered in two different areas: random matrix theory and
soliton equations.

Random matrix theory is many times called a new kind of statistical
mechanics, where one works with a ensemble of Hamiltonians satisfying
certain properties. It was born in physics in the attempts to analyze
the complicated spectra of heavy nuclei, where the phenomenom of
repulsion of energy levels is apparent. Its range of application has
been expanded intensively in many directions, like number theory,
quantum chaos and integrable systems. The applications to these two
later areas will be important for this work.

Quantum chaos can be roughly defined as the analysis of the  behavior
of classical chaotic systems when quantized. In other words, the
analysis of the Schr{\"o}dinger equation for a potential wich produces
chaotic behavior classicaly. Some researchers dislike the name quantum
chaos with the allegation that the Schr{\"o}dinger equation is linear,
then it cannot produce chaotic behavior. I think this is not a
consistent criticizing, for there is a fundamental difference between
finite and infinite dimensional dynamical systems: it is clear that a finite
dimensional linear system is integrable, i.e, it possesses a quantity
of integrals of motion equal to its dimension (number of degrees of
freedom), but the Schr{\"o}dinger equation does not possesses infinite
conservation laws, i.e, infinite densities commuting with the
Hamiltonian, except for some few potentials. This is a problem of
language, as already emphasized in \cite{cook}
To make this point clearer, some facts will be recalled. Suppose the
dynamical law
\begin{equation}
\frac{dX}{dt}=[H,X],
\end{equation}
where $X$ is some dynamical variable, like a function on classical
phase space, and the symbol $[,]$ denotes, as usual, the commutator
(Poisson bracket, Lie bracket...). Suppose some structure that enable
one to talk about adjoint (transpose, hermitean conjugate...) and that
$H$ is self-adjoint (symmetric, hermitian...), then it
is readly seen that
\begin{equation}
[H,X]=\lambda X,
\end{equation}
implies
\begin{equation}
[H,XX^*]=0,
\end{equation}
where the symbol $^*$ denotes adjoint. One can think in the harmonic
oscillator as an example, but this simple statement is
general. Instead of eigenfunctions, one can think of conserved
quantities, and this latter concept is in my sense more natural and
physically intellegible. It is important to insist in this point. The
eigenvalues of the Hamiltonian operator are given by
\begin{equation}
\lambda_{n}=\left<\Psi_n|H|\Psi_n\right>.
\end{equation}
Tough this definition is equivalent to usual one
$H|\left.\Psi_{n}\right>=\lambda_{n}|\left.\Psi_{n}\right>$ for time-independent
potentials, it is well suitable for time-dependent potentials and
makes clear the density character of the eigenvalues. With all this in
mind, another important connection can be made: for one dimension
only, what is the potentials wich give discrete time-independent
eigenvalues? With appropriate boundary conditions, this potentials are
the solutions of the KdV equations for shallow water waves. In
\cite{kruskal}
the appearance of the Schr{\"o}dinger operator is considered as merely
a coincidence. In fact, other nonlinear equations possessing soliton
solutions like mKdV and nonlinear Schr{\"o}dinger possesses as
Hamiltonian densities, this kind of Schr{\"o}dinger densities, altough
the Hamiltonian (symplectic) structure is not ever the same. To make
all these notions clearer, two simple systems will be used: the free
particle and the harmonic oscillator.

The pressurelles version of the NS equations is known by Burgers
equation. In fact, for more than one dimension it is called
generalized Burgers equation. In the Schr{\"o}dinger equation this is
simply the free-particle, with the well known propagator. It is
important to notice that the propagator is something that maps
densities. If one knows the particle is in some position at an initial
time, the initial density is an delta function. The propagator gives
the probability this particle stay at some later time $t$ in an
arbitrary position $x$. This propagator is the same as if one
considered the Brownian motion of this particle, with the Planck's
constant making the role of the viscosity coefficient. This can led
one to wonder if the Planck's constant is some kind of fundamental
viscosity coefficient. This is also a fact pointing to the connection
between dissipation and quantization.

Now some analysis of the harmonic oscillator. It is readly seen that
the abstract construction of the eigenvectors for an Hamiltonian, in
the case of the harmonic oscillator, these eigenvectors are given by
the so-called step operators $a, \dagger a$. Backing to the
geometrical side, it is instructive to consider these operators as
covariant derivatives. It is immediate that the invariant density is
the Gaussian function. More interesting is that if one tries to
calculate the ``curvature'', we get the curvature of the circle, wich
is in fact the harmonic oscillator topologically.

This work is an attempt to show that research in fluid dynamics,
quantum mechanics and quantum chaos can mutually benefit from each
other if a commom language is used.

We would like to Capes for financial support.

\bibliographystyle{prsty}
\bibliography{}

\end{document}